\documentclass[12pt, a4paper]{article}
\usepackage{amsfonts}
\usepackage{amsmath}
\usepackage{latexsym}
\usepackage[dvips]{graphics}
\usepackage{pgf}
\usepackage{hyperref}
\usepackage{pdflscape}
\usepackage{multicol}
\usepackage[german,english]{babel}
\usepackage{authblk}
\usepackage[onehalfspacing]{setspace}
\usepackage{lscape}
\usepackage{natbib}
\usepackage[nolist]{acronym}

\begin{document}

\title{\Large New results on the asymptotic and finite sample properties of the MaCML approach to multinomial probit model estimation}

\author[a]{Manuel Batram\footnote{Corresponding author; E-mail adresses: Manuel.Batram@uni-bielefeld.de, Dietmar.Bauer@uni-bielefeld.de}}
\author[a]{Dietmar Bauer}

\affil[a]{\footnotesize{Department of Economics, Bielefeld University, Postfach 10 01 31, D-33501 Bielefeld, Germany.}}
\date{}

\maketitle

\begin{acronym}
\acro{AIC}{Akaike Information Criterion}
\acro{APB}{Average Percentage Bias}
\acro{ASC}{Alternative Specific Constant}
\acro{BIC}{Bayesian Information Criterion}
\acro{bME}{bivariate Mendell-Elston}
\acro{CCL}{Composite Conditional Likelihood}
\acro{CDF}{Cumulative Distribution Function}
\acro{CEF}{Conditional Expectation Function}
\acro{CL}{Composite Likelihood}
\acro{CLAIC}{Composite Likelihood Akaike Information Criterion}
\acro{CLBIC}{Composite Likelihood Bayesian Information Criterion}
\acro{CML}{composite marginal likelihood}
\acro{c.p.}{ceteris paribus}
\acro{DCM}{Discrete Choice Models}
\acro{DGP}{Data Generating Process}
\acro{DM}{Decision Maker}
\acro{GHK}{Geweke-Hajivassiliou-Keane}
\acro{IC}{Information Criteria}
\acro{iid}{independent, identically distributed}
\acro{KLD}{Kullback-Leibler Divergence}
\acro{LLN}{Law of Large Numbers}
\acro{MaCML}{maximum approximate composite marginal likelihood}
\acro{MAE}{mean absolute error}
\acro{MC}{Monte Carlo}
\acro{MCMC}{Markov Chain Monte Carlo}
\acro{ME}{Mendell-Elston}
\acro{ML}{maximum likelihood} 
\acro{MNP}{multinomial probit}
\acro{MSE}{mean squared error}
\acro{MSL}{maximum simulated likelihood}
\acro{MVNCDF}{multivariate normal cumulative distribution function}
\acro{OLS}{Ordinary Least Squares}
\acro{PQD}{Positive Quadrant Dependence}
\acro{RMSE}{root mean squared error}
\acro{RUM}{random utility model}
\acro{RV}{Random Variable}
\acro{SEM}{Structural Equation Models}
\acro{SJ}{Solow-Joe}
\acro{SLLN}{Strong Law of Large Numbers}
\acro{ULLN}{Uniform Law of Large Numbers}
\acro{WLLN}{Weak Law of Large Numbers}
\acro{WTP}{Willingness to Pay}
 \end{acronym}

\begin{abstract}
\small{In this paper the properties of the maximum approximate composite marginal likelihood (MaCML) approach to the estimation of multinomial probit models (MNP) proposed by Chandra Bhat and coworkers is investigated in finite samples as well as with respect to asymptotic properties. Using a small illustration example it is proven that the approach does not necessarily lead to consistent estimators for four different types of approximation of the Gaussian cumulative distribution function (including the Solow-Joe approach proposed by Bhat). It is shown that the bias of parameter estimates can be substantial (while typically it is small) and the bias in the corresponding implied probabilities is small but non-negligible. Furthermore in finite sample it is demonstrated by simulation that between two versions of the Solow-Joe method and two versions of the Mendell-Elston approximation no method dominates the others in terms of accuracy and numerical speed. Moreover the system to be estimated, the ordering of the components in the approximation method and even the tolerance used for stopping the numerical optimization routine all have an influence on the relative performance of the procedures corresponding to the various approximation methods. Jointly the paper thus points towards eminent research needs in order to decide on the method to use for a particular estimation problem at hand.}
\vspace{1cm}
\end{abstract}


\section{Introduction}
Discrete choice models  are routinely used for modeling mode choice, 
destination choice, choice of travel time, route choice, vehicle purchase decision, activity choice and many other areas. They are used 
for example for the evaluation of new mode options, the design of pricing schemes for public transport, the adoption of new vehicle technologies,
the impact of the provision of travel time information as well as in the simulation of transportation systems. 
Discrete choice models have been used in cross sectional data sets based on revealed preferences as well as panel data sets combining 
revealed and stated preference data. 

Most commonly discrete choice models are formulated using the \ac{RUM} paradigm with the two basic models being the 
multinomial logit (MNL) and the \ac{MNP} model. While the MNL models suffers from the IIA assumption that led to the formulation
of 'generalized extreme value' (GEV) models \cite[see for example][]{train2009}  the \ac{MNP} model family offers better modeling flexibility at the expense of higher computational  costs. 

A major reason for the high computational costs is that the probit likelihood by definition involves a \ac{MVNCDF} which is analytically intractable and hence it is necessary to rely on approximation methods in order to evaluate the likelihood. 
For standard quadrature methods the relationship between the dimension of the integral and the computational complexity is of exponential order which renders those methods too time consuming for all but the smallest choice sets.

Therefore, the integral is usually approximated by \ac{MC}
simulations, which are in comparison less accurate. When combined with maximum
likelihood estimation those methods are known as \ac{MSL} approach. 
The most widely used method is the algorithm by \ac{GHK} \cite[see][p. 115]{train2009}. The assessment of integrals by simulation is computationally efficient because the
computational complexity of the simulation is an approximately linear function of the dimension of the integral \cite[see][p. 88f]{hajivassliliou2000}.
Like \ac{MC} methods in general \ac{MSL} is justified asymptotically, which induces the need to rely on many simulation runs and, nevertheless, leads to biased estimates whenever the number of \ac{MC} replications is finite and \cite[see][p. 250ff]{train2009}. 

In this regard, several authors have suggested that analytic approximations might
have the potential to offer a faster way to estimate \ac{MNP} models.
In fact, the use of analytic approximations -- namely
the Clark approximation -- to estimate \ac{MNP} models
predated the introduction of \ac{MSL} 
\citep{daganzo1977}. However, with the advent of
\ac{MSL} this approximation was deemed to be too imprecise and, therefore, its
importance vanished \citep{horowitz1982}. In general analytic
approximations of the  \ac{MVNCDF} are known to be less accurate than quadrature and simulation methods  \citep{joe1995} but they share neither the infeasibility of quadrature methods nor the long computation times of \ac{MSL}.

Utilizing an analytic approximation \citet{bhat2011} recently introduced the
\ac{MaCML} approach for  simulation-free estimation of \ac{MNP} models combining the
\ac{SJ}-approximation for the \ac{MVNCDF} and \ac{CML} estimation with the
specific aim to speed up the estimation of complex \ac{MNP} models. 

In a first simulation study \ac{MaCML} is reported to be up to 350 times faster than \ac{MSL} estimation while the accuracy of parameter recovery was at least at par with the latter \citep{bhat2011a}. 
A later simulation study revealed that the difference in computation times shrinks once \ac{MSL} estimation is also performed within the \ac{CML} framework but that there is still a reasonable performance gain which is then fully attributable to the analytic approximation  \citep{cherchi2016}. Furthermore, \citep{cherchi2016} report that \ac{MaCML} is faster and more accurate than several competing estimation procedures including two \ac{GHK} variants as well as Bayesian \ac{MCMC} estimation.

Despite those first results and the effort undertaken
by  \cite{bhat2011}  to provide theoretical justification for this specific
combination of methods, there are some open questions that warrant further
investigation: Firstly, we are not aware of any results addressing the
consistency of the estimator. The consistency of \ac{CML}
methods is rather well understood  \cite[see][]{varin2011} but it is unclear how
the \ac{MVNCDF} approximation interferes with estimation.

As the choice probabilities enter the log-likelihood in a nonlinear manner it is not clear that unbiased estimation of the probabilities leads to unbiased maximum (composite) likelihood estimators. 

Secondly, a
recent contribution by \cite{connors2014} notes that \citep{bhat2011} does not discuss the merits of alternative
analytic approximations of the \ac{MVNCDF}. They provide evidence that
the \ac{ME}-approximation is superior to the \ac{SJ}-approximation with regard to
accuracy and computation time. Those results were generated for a large set of
different \ac{MVNCDF}s and the approximations were judged by the ability to
replicate the \ac{MVNCDF}-probability from known parameters \cite{connors2014}.
Furthermore, \citep{trinh2015} have recently presented an improved variant of the \ac{ME}-approximation. Those results raise the question whether a combination of the \ac{ME}-approximation with \ac{CML} estimation is better suited to estimate \ac{MNP} models than the current \ac{MaCML} approach.

Note in this respect that it is common practice (see \cite{horowitz1982} and
\cite{kamakura1989}) to differentiate between (a) prediction accuracy, that is 
the predictive performance of the approximation with known parameters as assessed for example by \citep{connors2014}, 
and (b) estimation accuracy for the parameters themselves. 
 
It is
important to acknowledge that those two accuracy types are 
related but distinct concepts.  
This becomes even more important for some applications of \ac{MNP} models e.g. value of travel time or willingness to pay analysis where 
transformations of the parameter estimates (such as quotients of coefficients) 
and not the probabilities are the result primary of concern \cite[see][]{calfee2001}.

The papers \citep{bhat2011a} and \citep{cherchi2016} assess
the estimation accuracy of the \ac{SJ}-approximation for various (simulated)
data sets and \citep{kamakura1989} does the same for the
\ac{ME}-approximation,\footnote{\citep{kamakura1989} compares the
estimation and prediction accuracy of the \ac{ME}-approximation with
the Clark- and separate split approximation, see \citep{langdon1984}. It has been
shown repeatedly that those approximations are inferior (in terms of both accuracy concepts) to the
\ac{ME} (see \cite{kamakura1989}, \cite{rosa2003}) and,
therefore, those approximation are not part of our comparison.} but there is no
published comparison regarding the estimation accuracy of the \ac{SJ}- and
\ac{ME}-approximations in the context of \ac{MNP} model estimation.

The main contributions of this paper can be summarized as follows: First, we investigate the asymptotic bias of 
various \ac{MaCML} estimators using different approximations to the \ac{MVNCDF} in a simple example showing that 
the \ac{MaCML} approach does not deliver consistent estimators. By differentiation between predictive and estimation accuracy 
we demonstrate that the bias in estimated parameters can be sizeable while the bias in predicting choice probabilities 
in the example is small albeit not negligible.
Second, we compare the bias for different data generating processes in an attempt to identify situations where one of the approximations works better than the others. 
Third, we assess the
finite sample estimation accuracy of all methods using a cross-sectional \ac{MNP} model taken from \citep{bhat2011a}. These results are of
particular importance because they shed light on the typical computation time
of each estimator. It will be shown that there is no clear cut winner in those comparisons. 

The outline of the paper is as follows: In section~\ref{chap:probit} the model used for demonstration purposes is introduced. 
Section ~\ref{chap:app} describes the various approximation concepts used and surveys the literature on properties of the various concepts. Section~\ref{chap:results} then presents and discusses our findings and section~\ref{chap:concl} concludes the paper.

\section{The multinomial probit model} 
\label{chap:probit}
In this paper for illustration purposes we will use the following simple \ac{MNP} model based on the random utility function with an \ac{ASC} as the only explanatory variable, that is
\begin{equation*}
U_{nj} = \overline{b}_j + \epsilon_{nj} \qquad j = 1, \dots, J, \quad  n = 1, \dots N,
\end{equation*}
where at total of $N$ decisions $y_n \in \{1, \dots, 4\}$ between $J=4$ alternatives are observed. 

The
vector $\epsilon_{n} = (\epsilon_{nj})_{j=1,...,4}$ 
is assumed to be independently identically (i.i.d) normally distributed with mean vector of zero and variance matrix $\overline{\Sigma}$ assumed to be known. Therefore, we only need to fix $\overline{b}_1 = 1$ in order to ensure that the model is identified. Thus only three \ac{ASC}s constitute parameters to be estimated. 

Denote the difference between the utility induced by choosing the $j$-th and the $i$-th alternative by
$\tilde{U}^i_{nj} = U_{nj} - U_{ni}$ and define the $(J-1) \times J$ dimensional differencing matrix $\Delta_i$ obtained from
inserting a column of entries equaling $-1$ as the $i$-th column of the $J$ dimensional identity matrix.
Then the probability that individual $n$ chooses alternative $i$ is
defined as (defining $\overline{\textbf{b}}$ as the vector composed of all
$\overline{b}_j$s)
\begin{flalign}
\label{eq:ll-contri}
P_{i}(\overline{\textbf{b}}) = \mathbb{P}( y_n = i ) = \mathbb{P}(\tilde{U}^i_{nj} < 0 \quad  \forall j \not= i) 
= \Phi_3(\Delta_i \overline{\textbf{b}};
\textbf{0}, \Delta_i \overline{\Sigma} \Delta_i')
\end{flalign}
where  $\Phi_3(\cdot)$ denotes the distribution function of a
three-dimensional normal \ac{RV}. 

Note that the choice probabilities in this situation are a function of $\overline{\textbf{b}}$ and therefore the 
scaled log-likelihood $ll_N(\overline{\textbf{b}})$ under the assumption of independent choices can be written as 

\begin{flalign}
\label{eq:ll-probit}
ll_N( \overline{\textbf{b}}) = \frac{1}{N} \sum_{n=1}^{N} \sum_{j=1}^{J}
\mathbb{I}(y_n=j) \log P_j( \overline{\textbf{b}}) = \sum_{j=1}^{J}
\frac{N_j}{N} \log P_j( \overline{\textbf{b}})
\end{flalign}

where $\mathbb{I}(\cdot)$ denotes the indicator function and $N_j$ the number of individuals choosing alternative $j$. 

The resulting maximum likelihood estimator is defined as $\arg \max_{  \overline{\textbf{b}}} ll_{N} (  \overline{\textbf{b}})$. Under the assumption of independent observations this scaled log-likelihood converges
to a non-stochastic limiting function by the \ac{WLLN},

\begin{flalign}
\label{eq:conv}
ll_N( \overline{\textbf{b}})  \overset{P}{\rightarrow}\sum_{j=1}^{J}
P_j(\overline{\textbf{b}}^0) \log P_j( \overline{\textbf{b}}) = ll_0( \overline{\textbf{b}})
= \mathbb{E}[ ll_{N}(\overline{\textbf{b}})].
\end{flalign} 

Under the assumption that $\overline{\textbf{b}} \in B$ and $B$ is compact
this result immediately extends to uniform convergence 
(as in this case $0 < \inf_{\overline{\textbf{b}} \in B} P_j(\overline{\textbf{b}}) 
\le \sup_{\overline{\textbf{b}} \in B} P_j(\overline{\textbf{b}})  < 1, j=1,...,4$)
and therefore, subject to identification, to the consistency of the
maximum likelihood estimator \cite[see for example ][p. 114]{ferguson1996}. If in the likelihood the probability $P_j(\overline{\textbf{b}})$ is replaced by a continuous approximation $\hat P_j^{(m)}(\overline{\textbf{b}})$ it follows along the same lines that

\begin{flalign}
ll_N^{(m)}( \overline{\textbf{b}})  \overset{P}{\rightarrow}\sum_{j=1}^{J}
P_j(\overline{\textbf{b}}^0) \log \hat P_j^{(m)}( \overline{\textbf{b}}) = ll_0^{(m)}( \overline{\textbf{b}})
\end{flalign}

Again the corresponding maximizers converge to the maximizer of the limiting function $ll_0^{(m)}$. 
This allows the investigation of the asymptotic bias by examining the function $ll_0^{(m)}$ for different approximation 
approaches. 

\section{Methods for the approximation of choice-probabilities}
\label{chap:app}

This section  provides a total of four approximation methods for the \ac{MVNCDF} 
$\Phi_3(\textbf{b};
\textbf{0}, \textbf{R})$. As for diagonal transformation matrices $D$ it follows that 
$\Phi_3(D\textbf{b};\textbf{0}, D\textbf{R} D)$ we will in the following 
without restriction of generality assume that the coordinates have been scaled such that $\textbf{R}=[\rho_{ij}] \in \mathrm{R}^{3 \times 3}$ is 
a correlation matrix. 

All approximation methods are based on the \ac{SJ} and the \ac{ME} approach. 
The presentation is
short and focused on the \ac{MNP} model introduced in section~\ref{chap:probit}. For a more general discussion of the \ac{SJ}-approximation the reader is referred to \citep{joe1995}.
\citep{mendell1974} as well as \citep{kamakura1989} provide further information
regarding the \ac{ME}-approximation.

\subsection{Solow-Joe approximation}
\label{chap:SJ}
 
In the Solow-Joe approximation \citep{solow1990,joe1995} used by \citep{bhat2011} the multivariate normal
distribution is factorized into a product of conditional distributions, which
are in turn approximated by linear projections  \cite[see][p. 958]{joe1995}.\footnote{This is the first of two different approximations proposed by \citep{joe1995}. The second approximation, sometimes labeled as the  second-order approximation, is in theory more precise but its computation is more costly due to the appearance of tri- and quadvariate \ac{MVNCDF}s. \citep{sidharthan2012}  compared the influence both approximations have on parameter estimation and advises (except for rare special cases) against the use of the second-order approximation \cite[see][p. 134ff]{sidharthan2012}.}
For a three dimensional case of calculating the \ac{MVNCDF} for $X_j \le b_j, j=1,2,3$ 
where $X_j$ are standard normally distributed we obtain (using $I_j = \mathbb{I} (X_j \le b_j)$): 

\begin{flalign}
\Phi_3( {\textbf{b}};
\textbf{0}, \textbf{R}) &=
\mathbb{P}(X_{1} \le  {b}_{1}, X_{2} \le  {b}_{2}) \mathbb{P}(X_{3} \le  {b}_{3} |
X_{1} \le  {b}_{1},X_{2} \le  {b}_{2}) \label{eq:SJ}\\
& = \Phi_2( {b}_{1},  {b}_{2};\rho_{12}) \mathbb{E}[I_3 | 
I_{1} = 1, I_{2} =1]\nonumber\\ 
& \approx \Phi_2( {b}_{1},
 {b}_{2};\rho_{12}) \hat{p}^{3|12}( {\textbf{b}}, \textbf{R}) \nonumber \\ &=:
\hat{P}_4^{SJ:3|12}( {\textbf{b}}, \textbf{R}). \nonumber
\end{flalign}

Here the approximation replaces the conditional expectation by the linear projection 
\begin{equation}
\label{eq:linpro}
\hat{p}^{3|12}( {\textbf{b}}, \textbf{R}) := \Phi( {b}_{3}) + \textbf{q}
\textbf{Q}^{-1}[1- \Phi( {b}_{1}),1-
\Phi( {b}_{2})]',
\end{equation}
where $\textbf{Q}$ is a $2 \times 2$ matrix whose $(j,k)$s entry is
$Cov(I_j, I_k)$ where $j,k =
\{1,2\}$ and $\textbf{q}$ is a row vector with entries: $Cov(I_3, I_j)$, where
$j = \{1,2\}$. Note that for $i \not = j$ (with $\rho_{i,j}$ denoting the correlation between $X_i$ and $X_j$),

\begin{flalign*}
 Cov(I_i, I_j) &=
 \mathbb{E}[\mathbb{I}(X_i \le b_i) \mathbb{I}(X_j \le b_j)] - 
 \mathbb{E}[\mathbb{I}(X_i \le b_i)]  \mathbb{E}[\mathbb{I}(X_j \le b_j)]\\
 &= \Phi_2(b_i, b_j; \rho_{ij}) - \Phi(b_i)\Phi(b_j)
\end{flalign*}
and
\begin{equation*}
\label{eq:var}
 Cov(I_i, I_i) =
 \mathbb{V}(I_i) = \Phi(b_i)(1 - \Phi(b_i)).
\end{equation*}
 
 Therefore, the covariance matrix in ~\eqref{eq:linpro} is of the
 following form,

\begin{equation}
\label{eq:Rmatrix}
\textbf{Q} =
\begin{bmatrix}
\mathbb{V}(I_1) &  Cov(I_1, I_2) \\
Cov(I_1, I_2)  &\mathbb{V}(I_2)  \\
\end{bmatrix}.
\end{equation}

The approximation error of the \ac{SJ} approximation arises because in general 
$\hat{p}^{3|12}( {\textbf{b}}, \textbf{R}) \ne \mathbb{E}[I_3 | I_{1} = 1, I_{2} =1]$.

From the standpoint of
application, there are three issues that warrant further discussion.
Firstly, it should be clear that the ordering of the components in equation~\eqref{eq:SJ} is
arbitrary because the factorization would be equally valid for any permutation of
the components.
The major contribution of \citep{joe1995} over the older paper by \citep{solow1990} is the
suggestion to tackle this problem by computing the average over all possible
factorizations.
For larger choice sets this total enumeration is infeasible and \citep{joe1995} suggests to take the average of $10^2$ to $10^4$ randomly selected permutations \cite[see][p. 958]{joe1995}. However, \citep{bhat2011}
proposes that for the purpose of \ac{MaCML} estimation it is sufficient to use
only one randomly drawn permutation per estimation step 
\cite[see][p. 926f]{bhat2011}. In the respective simulations the performance improvement for utilizing two instead of one permutation is reported to be only marginal  \cite[see][]{bhat2011a}.
\citep{connors2014} assess the influence of the number of reorderings on the
prediction accuracy of the \ac{SJ}-approximation and conclude that averaging over 10 permutations provides a good compromise between computation time and
accuracy \cite[see][p. 127f]{connors2014}.

 Secondly, the linear projection \eqref{eq:linpro} involves the inverse
 of the matrix $\textbf{Q}$ and, therefore, the approximation is only compute-able
 when $\textbf{Q}$ is non-singular. Numerically $\textbf{Q}$ is almost singular if $|b_1|$ or $|b_2|$ is large or $b_1 \approx b_2$ and 
 $\rho_{ij} \approx 1$. Therefore numerical implementation needs to regularize in these situations. 

Thirdly, the approximation might yield values smaller that zero or larger than one 
which is problematic for calculating the log-likelihood which contains the term
\begin{flalign*}
 \log(\hat{P}^{SJ:3|1,2}( {\textbf{b}}, \textbf{R})) &=
 \log(\Phi_2( {b}_{1},  {b}_{2}) (\hat{p}^{3|1,2}( {\textbf{b}}, \textbf{R}))\\
 &=
 \log(\Phi_2( {b}_{1},  {b}_{2})) +
 \log(\hat{p}^{3|1,2}( {\textbf{b}}, \textbf{R}))
\end{flalign*}

The linear projection in the second term is not bounded between zero and one.
That positivity is not ensured proves to be particularly problematic because
in this case it is impossible to compute the approximation. This problem is 
discussed neither in \citep{joe1995} nor \citep{bhat2011}. However,
\citep{bhat2011a} briefly discusses the issue and state that it happens mainly
during gradient based optimization. Their solution for this
problem, which they describe as relatively rare, is to compute the
approximation based on a different permutation until the result is positive.
Likewise, the averaging suggested by Joe should ease this problem.

Finally $\hat{P}^{SJ:3|1,2}( {\textbf{b}}, \textbf{R})>1$ typically does not pose large problems as in this case $\log (\hat{P}^{SJ:3|1,2}( {\textbf{b}}, \textbf{R}))$ still is defined and the risk that the whole likelihood becomes larger than one 
usually is small. In any case this does not interfere with the maximization of the likelihood. 

\subsection{Mendell-Elston approximation}

The basic idea of the \ac{ME}
approximation as introduced by \cite{mendell1974} is to approximate the truncated normal by a normal distribution
with matching moments. For reasons of uniformity of presentation 
we will base the following exposition on the same factorization as \ac{SJ}
(see (\ref{eq:SJ})),

\begin{flalign}
\Phi_3( {\textbf{b}}; \textbf{0}, \textbf{R})
& = \mathbb{P}(X_1 \le  {b}_{1}) \mathbb{P}(X_2 \le  {b}_{2} |
X_1 \le  {b}_{1}) \mathbb{P}(X_{3}\le  {b}_{3} | X_{1}\le
 {b}_{1}, X_{2}\le  {b}_{2})  \label{eq:ME2}\\
& \approx \Phi( {b}_{1}) \Phi\left(\frac{ {b}_{2} - \mathbb{E}[X_2
\le  {b}_{2} | X_1 \le  {b}_{1}]}{\sqrt{\mathbb{V}[X_2 \le
 {b}_{2} | X_1 \le  {b}_{1}]}}\right) \Phi \left(\frac{ {b}_{3} -
\mathbb{E}[X_{3}\le  {b}_{3} | X_{1}\le  {b}_{1}, X_{2}\le
 {b}_{2}]}{\sqrt{\mathbb{V}[X_{3}\le  {b}_{3} | X_{1}\le
 {b}_{1}, X_{2}\le  {b}_{2}]}}\right) \label{eq:ME}\\
& =:
\hat{P}_4^{ME:123}( {\textbf{b}}, \textbf{R}). \nonumber
\end{flalign}

Therefore, in order to apply the approximation it is necessary to know all
the conditional moments that show up in ~\eqref{eq:ME}.
\citep{kamakura1989} utilizes the \ac{ME} approximation for \ac{MNP} model estimation and provides an algorithm for the calculation of the moments of
a truncated normal \cite[see][p. 256]{kamakura1989}. 
Using our notation an algorithm can be defined for the general case of 
$\Phi_K( {b}_{1},  ... ,  {b}_{K};\textbf{0}, \textbf{R})$
as follows: 
For $k=0,\dots,K$
and $l=0,\dots,k$ we start with $z_{k| 0}:= {b}_{k}$ and $r_{kl|
0}:=\rho_{kl}$. Then the algorithm for ME is given by
\begin{align}
	z_{k | l} &= \frac{z_{k| l-1}+a_{l| l-1}r_{kl|
	 l-1}}{\sigma_{k|l}} \label{eq:upper} \\ \nonumber
	a_{l | l-1} &= \frac{\phi(z_{l| l-1})}{\Phi(z_{l|
	l-1})} \\ \nonumber
	 r_{mo| l} &= \frac{r_{mo| l-1}-r_{ml|
	 l-1}r_{ol| l-1}a_{l|l-1}(a_{l| l-1}+z_{l|
	 l-1})}{\sigma_{m|l} \sigma_{o|l}} \\ \nonumber
	 \sigma_{k|l} &= 1-r_{kl| l-1}^{2}a_{l| l-1}(a_{l|
	l-1}+z_{l| l-1})
\end{align}

where $m=1,\dots,k$ and $o=1,\dots,k$ with $m\neq o$. Furthermore, note that
$z_{l| l-1}$ denotes the short-form of $z_{l| l-1,\dots,1}$. In combining
the recursive algorithm with ~\eqref{eq:ME} we denote the general \ac{ME}-approximation as,

\begin{align}
\label{eq:ME_f}
	\Phi_K( {b}_{1},\dots, {b}_{K};\textbf{R})
	&=\mathbb{P}(X_{1}\le {b}_{1})\prod_{k=2}^{K}{\mathbb{P}(X_{k}\le {b}_{k}| X_{k-1}\le {b}_{k-1},\dots,X_{1}\le {b}_{1})} \\
	& \approx \Phi(z_{1|0})\cdot \Phi(z_{2|1})
	\cdot \dots \cdot \Phi(z_{K| K-1}). \nonumber
\end{align}

Again, the ordering of the components in the factorization in
~\eqref{eq:ME2} is arbitrarily chosen from the possible
permutations of the components. However, because of the very nature of the \ac{ME}-approximation
it is possible to tackle this problem without brute-force averaging. The idea is to
pick the factorization, whose factors are closest to
the normal distribution. 
In this regard, \citep{langdon1984} derived the first four moments of factors
resulting from different factorizations of a three alternative probit model. The skewness and kurtosis of those factors were closest to the normal distribution
whenever the terms were sorted with the standard deviation of the utility differences in ascending order. 

With his recursive algorithm in mind \citep{kamakura1989} presents a slightly modified
procedure. The first term is still the one with the minimum standard deviation
but all subsequent ones are chosen to have the minimal $\sigma_{i|j}$ amongst the
remaining integration limits.
The simulations by \citep{connors2014} reveal that instead of
only focusing on the standard deviations the optimal
factorization (in terms of approximation accuracy) uses a 
decreasing order of the upper integration limits.

\subsection{Bivariate Mendell Elston approximation} 
Recently, \cite{trinh2015} presented an improved version of the
\ac{ME}-approximation.  The major contribution is the use of bivariate normal
distributions within the \ac{ME}-approximation and the development  of formulas for the corresponding moments whose derivation is largely based on prior results from \citep{muthen1990}. Furthermore, their algorithm utilizes several pre-processing steps
routinely used in conjunction with quadrature methods (e.g. handling the covariance matrix in its Cholesky factorized form). However, their
approximation is still based on the general idea of \ac{ME} and basically results in the replacement of the univariate with bivariate
normal distributions in ~\eqref{eq:ME_f}. We will abbreviate this special variant of \ac{ME} as \ac{bME}.

In the simulations that accompany their theoretical work \citep{trinh2015} confirm the aforementioned results of \citep{connors2014} and show that reordering leads to considerable gains with regard to the predictive accuracy. Furthermore, they illustrate that
for their bivariate \ac{ME}-approximation it is possible to perform the
reordering based on the results from the univariate \ac{ME}, which provides
substantial benefits regarding computation time.
\\
 
\subsection{Comparison of \ac{MVNCDF} approximations: literature review}
\label{chap:comp}

In this section we briefly review the literature on comparisons between the \ac{SJ}- and \ac{ME}-approximation with regard computation time and predictive accuracy. From the previous paragraphs we know that both
approximations rely on the availability of a method to compute one dimensional
\ac{MVNCDF}s. For the computation of the \ac{SJ}
and the \ac{bME} approximation additional two dimensional \ac{MVNCDF}-integrals
need to be solved, which is computationally more burdensome.\footnote{A comparison of the computational burden of 
evaluating the one dimensional Gaussian cdf versus a bivariate \ac{MVNCDF} in MATLAB led to roughly a factor 15.}
Furthermore, the computational costs of the two strategies to improve the
approximation accuracy by averaging over different permutations of the components 
vary by a large margin. \citep{joe1995} proposes that
averaging over all alternative is the optimal strategy for the
\ac{SJ}-approximation, however, this strategy has fixed computational costs of $O(K!)$. 
Optimal reordering
strategies
for the \ac{ME}-approximations boil down to a sorting operation with worst case costs of $O(K^2)$.

Those preliminary observations are reflected in the empirical findings regarding
the comparison of computation times. 
\citep{joe1995} states (without providing detailed quantitative evidence) that the \ac{ME}-approximation 
is faster than
the all-permutations \ac{SJ}-approximation. He attributes this to the computational burden
imposed on the \ac{SJ}-approximation by the need to evaluate \ac{MVNCDF}-integrals of dimension higher than one \cite[see][p. 960]{joe1995}.

\citep{connors2014} compare the computation times of an optimally ordered \ac{ME}-approximation and
a \ac{SJ}-approximation with 10 reorderings  
in a broad set of possible covariance configurations. Depending on
the dimension of the \ac{MVNCDF} the \ac{SJ}-approximation is on average 12
to 63 times slower than the \ac{ME}-approximation  \cite[see][p. 130]{connors2014}.\footnote{It is worth noting
that the simulations by \citep{connors2014} reveal that as expected both
approximations are faster than the \ac{GHK}-algorithm and quadrature methods.
\citep{trinh2015} confirm parts of this finding for the univariate and bivariate
\ac{ME}-approximation, which are shown to be at least 10 times faster than
Monte Carlo evaluation of the integral.}
Similar results are reported by \citep{rosa2003}, who evaluates the
approximations in the context of Probit-Based Stochastic
User Equilibrium models and states that the mean computation time for the
\ac{SJ}-approximation is up to 100 times larger than the respective value of the \ac{ME}-approximation. Furthermore, the standard deviation of the computation time is higher for the
\ac{SJ}-approximation \cite[see][p. 181]{rosa2003}. \citep{rosa2003} utilizes averaging over all permutations whenever the integral dimension is smaller than 10 and 1000 permutations for larger integrals.
 Finally, \citep{trinh2015}
compare the computation times of \ac{SJ} and their bivariate
\ac{ME}-approximation for one specific five-dimensional \ac{MVNCDF} and find
that their algorithm is 16 times faster than a \ac{SJ}-approximation with
averaging over all permutations. 

The results regarding the prediction accuracy are not as clear-cut as the result
regarding computation time. On the one hand the results by \citep{joe1995}
suggest that the approximations are comparable.
Furthermore, \citep{trinh2015} also present simulations were the
\ac{SJ}-approximation with all permutations is superior to the univariate as
well as the bivariate \ac{ME}-approximation.

On the other hand the results of \citep{connors2014} and \citep{rosa2003} suggest that the \ac{SJ}-approximation is inferior with regard to almost any performance metric. \citep{rosa2003} reports that the average percentage error is always larger for the \ac{SJ}-approximation when compared to the \ac{ME}-approximation 
\cite[see][p. 178ff]{rosa2003}. The average percentage error of the
\ac{ME}-approximation is found to be up to 20 times lower than the error of the
\ac{SJ}-approximation. However, \citep{rosa2003} illustrates that it is possible
to narrow down the spread by excluding cases with small true
probabilities from the simulations. This is in line with \citep{trinh2015}, who
find that the predictive accuracy of both \ac{ME}-approximations is lower for smaller true
probabilities.

\citep{connors2014} provide a broad range of simulation settings. However, the
setup differs slightly from \citep{rosa2003} and \citep{joe1995} because the
accuracy/error calculations are done for the logarithm of the probabilities.
For the simulation setting of \citep{connors2014} the
\ac{ME}-approximation seems to outperform the \ac{SJ}-approximation for every
number of choice alternative regardless of the performance measure reported.

It is worth noting that the results of \citep{connors2014} might be
influenced by the method used to generate the covariance matrices. The
relationship between a correlation ($\textbf{R}$) and a covariance matrix
($\Sigma$) is,
\begin{equation}
\label{eq:covar}
\Sigma = \mathcal{D}(\textbf{s}) \textbf{R} \mathcal{D} (\textbf{s}),
\end{equation}
where $\mathcal{D}(\textbf{s})$ is a diagonal matrix of the same dimension as
\textbf{R}, whose entries are the standard deviations $s_j$. The covariance matrices in \citep{connors2014} are
computed as $\Sigma' = \alpha \mathcal{I} + \textbf{R}$, where
$\mathcal{I}$ is the identity matrix and $\alpha$ is a scalar and $\alpha$ as
well as the correlation matrix $\textbf{R}$ are generated at random. When compared to equation~\eqref{eq:covar} it is clear that by construction the methodology of \citep{connors2014} typically produces covariance matrices with small correlations for sizeable $\alpha$. 

However, the \ac{ME}-approximation works particularly well for \ac{MVNCDF} with small
correlations (see \citet[p. 455f]{rice1979} and \citet[p. 995]{trinh2015}) and,
therefore, this method might benefit from the experimental methodology.

In summing up this section we conclude that the literature 
indicates that the \ac{ME}-approximation is supposedly superior to the \ac{SJ}-approximation with regard to computation time as well as prediction accuracy. It is worth noting that most comparisons considered the \ac{SJ}-approximation with averaging over all permutations, while \citep{bhat2011} suggests to use only one permutation for \ac{MaCML} estimation. Furthermore, we outlined that the appearance of small true probabilities as well as the correlation structure might influence the performance of the approximations and we will devote special care to those cases in the following simulations.

\section{Main findings}
\label{chap:results}
In this section the properties of estimators obtained by maximizing the approximated composite marginal likelihood 
are discussed. Here estimation is performed using observations $\omega_N \in \Omega_N$ of sample size $N$ where 
$\omega_N$ accounts for all explained and explanatory variables. In the motivating example $\omega_N = [y_n]_{n = 1,...,N}$. 

The scaled logarithm of the 
composite likelihood\footnote{Note that (standard) maximum likelihood estimation is included in the \ac{CML} framework as one special case.} is denoted as $ll_N(\theta;\omega_N)$ where $\theta \in \Theta \subset \mathbb{R}^{M}$ 
denotes the parameter vector to be estimated. For the model in section~\ref{chap:probit} thus 
$\theta = \overline{\textbf b} \in \mathbb{R}^{4}$. Then the CML approach obtains the estimator 
$$
\hat \theta_N := \mbox{argmax}_{\theta \in \Theta} ll_N(\theta;\omega_N)
$$
Under well known appropriate regularity conditions we obtain $ll_N(\theta;\omega_N) \to ll_0(\theta)$ almost 
surely  both pointwise and uniformly in $\theta \in \Theta$ where the limiting criterion function $ll_0$ is continuous in $\theta$ and uniquely maximized at the true parameter vector $\theta_0$.

The \ac{MaCML} approach defines the pseudo CML 
$ll_N^{(m)}(\theta;\omega_N)$ by replacing the \ac{MVNCDF} $P_i(\theta;\omega_n)$ by the approximation $\hat P_i^{(m)}(\theta;\omega_n)$. Here $m\in \{ \mbox{SJ-A, SJ-1, ME, bME} \}$ in this paper. 
Consequently the \ac{MaCML} estimator is defined as 
$$
\hat \theta_N^{(m)} := \mbox{argmax}_{\theta \in \Theta} ll_N^{(m)}(\theta;\omega_N)
$$
Again $ll_N^{(m)}(\theta;\omega_N) \to ll_0^{(m)}(\theta)$ almost surely uniformly in combination with continuity of the limiting criterion function $ll_0^{(m)}(\theta)$ leads to convergence of $\hat \theta_N^{(m)}$ to 
$\theta_0^{(m)}$, the maximizer of $ll_0^{(m)}(\theta)$ over $\Theta$ (assuming uniqueness of the maximizer). 
It follows that the asymptotic bias is given by $\theta_0^{(m)}-\theta_0$ and hence can be investigated based 
solely on  $ll_0^{(m)}(\theta)$.

Therefore in this section, first we investigate the properties of $\theta_0^{(m)}-\theta_0$  (section~\ref{chap:large}) followed by results for the finite-sample case in section~\ref{chap:finite}. 
All calculations are done in MATLAB R2015b,\footnote{Some of our functions are derived from \citep{mvnxpb} as well as from the R and GAUSS code of \citep{bhat2011a}.} whose \textit{mvncdf} function is used to compute the 'true probabilities'. This function is stated to provide an
accuracy of $10^{-8}$ for integrals of dimension smaller than four and
$10^{-4}$ for integrals of higher dimension.

\subsection{Large-Sample Properties of \ac{MaCML} estimators}
\label{chap:large}

In this section we assess the consistency of \ac{MaCML} estimation with four different approximations,
\begin{enumerate}
    \item \ac{SJ} as an average over all permutations (SJ-A),
  \item \ac{SJ} with one random permutation (SJ-1),
  \item semi-optimally ordered, univariate \ac{ME} (ME),
  \item semi-optimally ordered,  bivariate \ac{ME} (\ac{bME}).
\end{enumerate}
The \ac{bME} variant we use combines \ac{bME} with a univariate reordering in order to speed up computations: 
For \ac{ME} we apply a reordering scheme where we reorder the \ac{ASC}s in descending order. Even though the \ac{ASC}s are a prominent part of the upper integration limits of the \ac{ME} approximations (see (\ref{eq:upper}) and remember that $z_{k| 0}:= {b}_{k}$ ) this is not the optimal reordering suggested by \citep{connors2014}. Most notably we do not reassess and 
potentially alter the reordering after each component of the approximation (see (\ref{eq:ME_f})) is computed. 
However, preliminary tests have shown that the optimal reordering scheme, which reorders after each component is 
computed, leads to discontinuities in the limiting likelihood function $ll_0^{(ME)}$. 
Clearly discontinuities pose severe threats for the typically applied gradient methods for numerical function
optimization. 

In order to apply SJ-1 in the context of the limiting distribution, which is the expected value of the scaled 
likelihood, we compute the mean of all possible log likelihood values resulting from the single ordering 
approximations. In comparison to SJ-A this mean value
is an outer mean (averaging the logarithms of the approximated probabilities), 
while in the case of SJ-A the mean is computed as an inner mean (first averaging, then taking logarithms) 
of the approximated probabilities.

We simulate limiting (pseudo-)likelihoods for  8.000 different models of the type introduced in section~\ref{chap:probit}. The \ac{ASC}s as well as the standard deviations are drawn from
uniform distributions $[-L,L]$ and $[0,K]$, respectively. We iterate through different bounds ($L \in \{1,2\}$,  $K \in \{1,2,3,4\}$)\footnote{Values of $L$ larger than 2 lead to an extensive share of models with at least one very small true probability (e.g. for K=1, L=4 we observed that 96 percent of the models have at least one true probability smaller than 0.001).} and simulate 1000 instances for each combination.

The correlation matrices are generated by the Vine-based method of \citep{lewandowski2009} which is designed to uniformly generate correlation matrices from the space of positive definite correlation matrices 
\cite[see][]{lewandowski2009}. In order to simulate matrices with small correlations
it is possible to depart from the uniform sampling and instead sample with
probability proportional to $\det|\textbf{R}|^{\eta - 1}$, where $\eta$ scales
the penalty for the appearance of large correlations. The respective covariance
matrices are calculated using ~\eqref{eq:covar}. 

We search the maximum of each limiting (pseudo-)likelihood using an optimization
algorithm, which is constrained to [-2L,2L] for each parameter and initialized at
the true values in order to speed up the calculations. As this setup essentially
imposes an additional condition of local identifiability around the true value the 
results presented below are best-case estimates of the asymptotic bias.

We report the errors as measured by a) the \ac{RMSE},
$\sqrt{\frac{1}{B}\sum_{b=1}^B \tau(\hat{\theta_b},\theta_b^0)^2}$ and b) the
 \ac{MAE}, $\frac{1}{B}\sum_{b=1}^B[\tau(\hat{\theta_b},\theta_b^0)]$, where $\tau(\hat{\theta_b},\theta_b^0)$
 denotes a transformation of the estimator $\hat{\theta_b}$ and the true parameter vector $\theta_b$ used in the
 $b$-th Monte Carlo replication. The mapping $\tau$ hereby corresponds either to the difference in a particular coordinate $\hat{\theta_b}_m-\theta_{b,m}^0$, the maximum over all coordinates
 $\max_{m=1,...,M}|\hat{\theta_b}_m-\theta_{b,m}^0|$
 or the maximal deviation in the corresponding probabilities 
 $\max_{j=1,...,J} |P_j(\hat \theta_b) - P_j(\theta_b^0)|$.

The first column of Table~\ref{tab:e_err_para} shows the average asymptotic estimation error of the approximations with regard to the estimated parameters. The first and by far most important observation is that even for a simple model all approximations yield inconsistent estimators. For reasons already discussed both \ac{SJ} variants and \ac{bME} are comparable accuracy-wise, while the univariate ME approximation has the worst accuracy. Note, that the better performance of approximations that utilize a bivariate \ac{MVNCDF} is expected because the 
computation of each true probability involves only the evaluation of a three dimensional \ac{MVNCDF} (see (\ref{eq:ll-contri})).

\begin{figure}%
{\caption{Empirical distribution of the absolute
errors for 8.000 models}\label{fig:ecdf}}%
\begin{tabular}{@{}r@{}} 
\includegraphics[width=\textwidth]{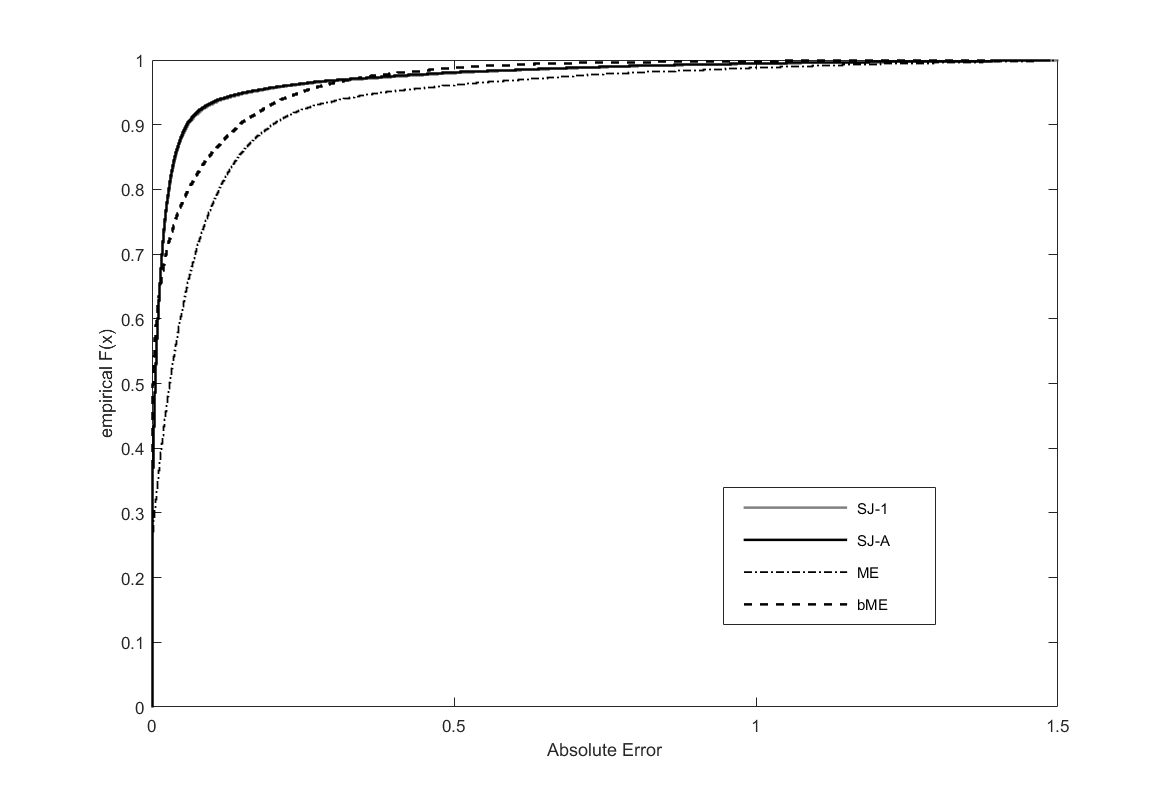}\\
\end{tabular}
\end{figure}

\begin{table}[t]
\centering
\caption{Parameter estimation error for various thresholds  $x$  regarding small
true probabilities ($\min P_j(\theta_0) \geq x$)}
\begin{tabular} 
 {l l *{4}{c} } 
& $x$ & 0  &  0.001 & 0.01 & 0.05\\ \hline 
max. RMSE & SJ-A & 0.196 &0.049 &0.029 &0.023 \\ 
& SJ-1 & 0.194 &0.040 &0.027 &0.023 \\  
 & ME & 0.331 &0.103 &0.091 &0.082 \\   
 & bME & 0.140 &0.158 &0.138 &0.113 \\  \hline 
max. MAE & SJ-A & 0.058 &0.021 &0.018 &0.015 \\ 
& SJ-1 &0.057 & 0.021 &0.017 &0.015 \\  
 & ME &0.143 & 0.069 &0.064 &0.061 \\  
 & bME & 0.063 &0.086 &0.082 &0.074 \\  \hline 
number of obs. &  & 8000 &5218 &4041 &2384 \\  
\hline 
 \end{tabular} 

\label{tab:e_err_para}

\end{table}

Furthermore, the \ac{SJ} variants show a large difference between \ac{MAE} and \ac{RMSE}, which hints to the fact that the \ac{SJ} might produce a higher share of large errors, especially when compared to the values for \ac{bME} which have a smaller difference. Figure~\ref{fig:ecdf}, which
depicts the empirical distribution functions of the maximum absolute error for
the 8.000 models of the base case, provides further evidence in this regard. As is clearly visible from the intersecting empirical cumulative density functions that neither \ac{bME} nor the SJ variants (whose graphs are practically indistinguishable) are uniform best. However, we conclude from the figure that \ac{ME} is dominated by the three other approximations with respect to the estimation accuracy measured by \ac{MAE}. Additionally, it is clearly visible that both \ac{SJ} variants and \ac{bME} have extremely small asymptotic errors for about 70 percent of the simulation instances, while this share for \ac{ME} is only 30 percent. Therefore, even though the \ac{MaCML} estimators are inconsistent, we can conclude that at least for the simple model considered here the asymptotic error is often very small.

In order to assess the effect that the occurrence of small true probabilities has
on the estimation Table~\ref{tab:e_err_para} includes results for subsets of the
simulation. Those subsets contain only models whose smallest true
probability is larger than the respective thresholds. The discussion from
section~\ref{chap:comp} suggests that the performance of the approximations
should improve with rising thresholds. We are able to observe this effect for all approximations. However, while the accuracy of \ac{bME} also improves the
change is far less substantial and, therefore, both \ac{SJ} variants outperform \ac{bME} for all but the base case.

For several applications not the coefficients but the predicted
probabilities at the estimated coefficients, which we will from now on call
estimated probabilities, are of primary concern.
Because the mapping from parameter estimates to probabilities is non-linear the
errors in estimated probabilities need to be assessed separately from the
estimation error of the parameters.
Table~\ref{tab:e_err_prob} shows that those errors are substantially smaller
than the errors in the estimated parameter. This might be mainly an
effect of the tail regions of \ac{MVNCDF} being insensitive to
changes in the parameters. It is, however, noteworthy that there are mixed
results for the performance gains resulting from the exclusion of small
probabilities. While both \ac{SJ} variants and \ac{ME} show small improvements the accuracy of  \ac{bME} decreases notably. When interpreting those results, it is important to keep in mind that the \ac{MaCML} approach is pitched for situations where \ac{MSL} is infeasible \cite[see][]{bhat2011a} because of large choice sets. With a large number of alternatives in a choice set the chance is high that some have a choice probability which is small.

\begin{table}[t]
\centering
\caption{Probability estimation error for for various thresholds  $x$  regarding small true probabilities ($\min P_j(\theta_0) \geq x$)}
\begin{tabular} 
 {l l *{4}{c} } 
& $x$ & 0  &  0.001 & 0.01 & 0.05\\ \hline 
max. RMSE & SJ-A & 0.004 &0.003 &0.003 &0.003 \\ 
& SJ-1 &   0.004 &0.003 &0.003 &0.003 \\  
 & ME  & 0.015 &0.014 &0.014 &0.013 \\   
 & bME  & 0.013 &0.016 &0.017 &0.020 \\  \hline 
max. MAE  & SJ-A &   0.002 &0.002 &0.002 &0.002 \\ 
& SJ-1 &0.002 & 0.002 &0.002 &0.002 \\  
 & ME  &0.012 & 0.011 &0.010 &0.010 \\  
 & bME  & 0.007 &0.010 &0.012 &0.015 \\  \hline 
number of obs. &  & 8000 &5218 &4041 &2384 \\  
\hline 
 \end{tabular} 

\label{tab:e_err_prob}
\end{table}

Furthermore we have investigated the interrelation between the magnitude of
correlation and estimation accuracy. The setup is similar to the simulations
used in the preceding paragraph but for every $\eta$ we only draw 4.000 true models (500 for each combination of $L$ and $K$). As described in the
introduction to this section increased values of $\eta$ lead to fading
correlations. The results are shown in table~\ref{tab:e_err_eta}. The first
column denotes the case of uniform sampling $\eta = 1$, which was used for the
previous simulations. Reading from left to right, it is clearly visible that as the
correlations get weaker (for increasing $\eta$) \ac{MAE} and \ac{RMSE} are declining. This observation,
which applies to all approximations is not surprising because a \ac{MVNCDF} without
correlations is just the product of its marginal distribution functions.

\begin{table}[t]
\centering
\caption{Parameter estimation error for different correlation structures (larger values of $\eta$ correspond to weaker correlations)}
\begin{tabular} 
 {l l *{5}{c} } 
&  $\eta$&   1 &  5 & 50&100  \\ \hline  
max. RMSE & SJ-A & 0.165 &0.134 &0.113&0.117  \\ 
& SJ & 0.171 &0.132 &0.116&0.115  \\  
 & ME & 0.270 &0.240 &0.228&0.227 \\   
 & bME & 0.131 &0.115 &0.104&0.103  \\  \hline 
max. MAE & SJ-A & 0.049 &0.036 &0.029&0.029  \\ 
& SJ &0.049 & 0.036 &0.029&0.028  \\  
 & ME &0.126 & 0.105 &0.097&0.096  \\  
 & bME & 0.060 &0.055 &0.048&0.048  \\  
\hline 
 \end{tabular} 

\label{tab:e_err_eta}
\end{table}

To sum up the large-sample results indicate that the \ac{SJ}
variants and \ac{bME} are at par with regard to the parameter estimation error.
\ac{SJ} is superior for situations without small probabilities and for estimated
probabilities. Furthermore, this section showed that with regard to the limiting
estimation error \ac{ME} is inferior to the other three approximation in almost any setting. The performance of all approximations
is improved almost equally as correlations get weaker. 

\subsection{Finite-Sample Properties of \ac{MaCML}
estimators}
\label{chap:finite}

We assess the finite sample properties of \ac{MaCML} estimation with three different approximations,
\begin{enumerate}
  \item \ac{SJ} with one random permutation (SJ-1),
  \item semi-optimally ordered, univariate \ac{ME} (ME),
  \item semi-optimally ordered,  bivariate \ac{ME} (\ac{bME}).
\end{enumerate}

The reordering strategies match those from the large-sample simulation but in this section we apply \ac{MaCML} estimation to a model selected from the literature (see \citet{bhat2011a} as well as \citet{cherchi2016}). The model is a cross-sectional mixed multinomial probit model with five alternatives. Those alternatives are explained by five explanatory variables (drawn from independent standard normal distributions) and their respective parameters are assumed to be an instance of a multivariate normal distribution with mean $\textbf{b} = (1.5, -1, 2, 1, -2)$ and covariance matrix,

\begin{equation*}
\Gamma =
\begin{bmatrix}
 1 & - 0.50 & 0.25 & 0.75 & 0\\
-0.50 & 1 & 0.25 & -0.50 & 0\\
0.25 & 0.25 & 1 &0.33 & 0\\
0.75 & - 0.50 & 0.33 & 1 & 0\\
0 & 0 & 0 & 0 & 1
\end{bmatrix}.
\end{equation*}

The error terms are assumed to be normally distributed with a mean of zero and a
diagonal covariance matrix whose entries are 0.5. We generated 20 data sets of size 5000 by drawing values of the vector $\textbf{b}_i$ and the error term from their respective distribution.  Based on those values we calculated the utility and the chosen alternative is the one with the highest utility.

Our aim is to estimate the linear ($\textbf{b}$) as well as the covariance 
parameters. The covariance parameters are estimated using the corresponding Cholesky
decomposition ($\Gamma = LL'$). In order to find the optimum of the likelihood function we rely on the BFGS-algorithm provided by MATLABs \textit{fminunc} function. 
To ensure competitive computation times we have derived the respective analytic gradients for all pseudo-likelihoods but other than that relied on the default options. Following \citep{bhat2011a} we initialize the optimizer at the true values and use one random permutation per observation for the \ac{SJ} approximation, which stays the same during the optimization. We report the mean estimate over all 20 data sets as well as the estimation errors in form of the mean \ac{MAE} over all 20 data sets for each respective parameter. Note that the \ac{MAE} is calculated as the mean of the respective  absolute errors for each data set and is not equivalent to the absolute difference between the mean estimate and the true value.\footnote{We do not consider the \ac{APB} reported in \citep{bhat2011a}, due to its known shortcomings. For reasons of comparability we calculated the \ac{APB} as described in \citep{bhat2011a}: For the \ac{SJ} approximation our simulations resulted in an \ac{APB} of 3.9\%, while \citep{bhat2011a} report an \ac{APB} of 5.5 \% for the same experimental setting.}

\begin{landscape}
\begin{table}[t]
\centering
\caption{Mean parameter estimates, (mean) absolute errors and standard deviations (SD) of those errors across 20 simulated data sets of size 5000 \vspace{.2cm}} 
\begin{tabular} 
 { l c *{3}{c} *{3}{c} *{3}{c}} 
Parameter & true value & \multicolumn{3}{c}{SJ-1} & \multicolumn{3}{c}{ME} & \multicolumn{3}{c}{bME}\\  
&  & {\small mean} &  {\small MAE} &  {\small SD MAE} & {\small mean}  & {\small MAE} &  {\small SD MAE} & {\small mean}  & {\small MAE} &  {\small SD MAE} \\ \hline 
\multicolumn{11}{l}{{\small linear parameters}}  \\ 
$b_1$ & 1.50 & 1.53 & 0.12 & (0.112) & 1.47 & 0.04 & (0.095) & 1.47 & 0.04 & (0.067)   \\ 
$b_2$ & -1.00 & -1.02 & 0.09 & (0.045) & -0.99 & 0.03 & (0.069) & -0.98 & 0.03 & (0.043)   \\ 
$b_3$ & 2.00 & 2.04 & 0.18 & (0.052) & 1.97 & 0.04 & (0.066) & 1.95 & 0.05 & (0.042)   \\ 
$b_4$ & 1.00 & 1.01 & 0.09 & (0.048) & 0.96 & 0.04 & (0.039) & 0.97 & 0.03 & (0.046)   \\ 
$b_5$ & -2.00 & -2.03 & 0.18 & (0.105) & -1.96 & 0.05 & (0.080) & -1.95 & 0.06 & (0.077)   \\ 
\multicolumn{11}{l}{{\small covariance parameters (Cholesky factors)}}  \\ 
$l_{11}$ & 1.00 & 1.03 & 0.10 & (0.053) & 1.05 & 0.06 & (0.059) & 1.03 & 0.04 & (0.045)   \\ 
$l_{21}$ & -0.50 & -0.52 & 0.07 & (0.044) & -0.51 & 0.06 & (0.054) & -0.52 & 0.06 & (0.058)   \\ 
$l_{31}$ & 0.25 & 0.26 & 0.05 & (0.081) & 0.14 & 0.11 & (0.078) & 0.31 & 0.07 & (0.073)   \\ 
$l_{41}$ & 0.75 & 0.75 & 0.09 & (0.138) & 0.75 & 0.08 & (0.127) & 0.73 & 0.06 & (0.059)   \\ 
$l_{51}$ & 0.00 & 0.00 & 0.07 & (0.056) & 0.16 & 0.16 & (0.107) & -0.07 & 0.08 & (0.041)   \\ 
$l_{22}$ & 0.87 & 0.90 & 0.09 & (0.046) & 0.97 & 0.11 & (0.062) & 0.87 & 0.04 & (0.036)   \\ 
$l_{32}$ & 0.43 & 0.41 & 0.08 & (0.053) & 0.45 & 0.05 & (0.075) & 0.38 & 0.08 & (0.044)   \\ 
$l_{42}$ & -0.14 & -0.18 & 0.09 & (0.069) & -0.18 & 0.08 & (0.044) & -0.18 & 0.08 & (0.045)   \\ 
$l_{52}$ & 0.00 & -0.02 & 0.09 & (0.049) & -0.01 & 0.08 & (0.063) & -0.01 & 0.08 & (0.052)   \\ 
$l_{33}$ & 0.87 & 0.86 & 0.11 & (0.053) & 0.85 & 0.06 & (0.071) & 0.86 & 0.06 & (0.046)   \\ 
$l_{43}$ & 0.24 & 0.27 & 0.08 & (0.071) & 0.30 & 0.10 & (0.076) & 0.25 & 0.07 & (0.029)   \\ 
$l_{53}$ & 0.00 & 0.01 & 0.09 & (0.047) & 0.19 & 0.20 & (0.058) & -0.06 & 0.10 & (0.049)   \\ 
$l_{44}$ & 0.60 & 0.59 & 0.08 & (0.044) & 0.65 & 0.09 & (0.029) & 0.59 & 0.06 & (0.054)   \\ 
$l_{54}$ & 0.00 & -0.01 & 0.10 & (0.124) & -0.06 & 0.09 & (0.057) & 0.01 & 0.08 & (0.060)   \\ 
$l_{55}$ & 1.00 & 0.98 & 0.11 & (0.047) & 0.91 & 0.10 & (0.042) & 0.98 & 0.05 & (0.048)   \\ 
\hline 
\multicolumn{2}{l}{{\small Mean MAE and mean SD across parameters}}  & \multicolumn{3}{c}{{0.097 (0.067)}} &  \multicolumn{3}{c}{{0.081 (0.067)}}   & \multicolumn{3}{c}{{0.061 (0.051)}}  \\  
\multicolumn{2}{l}{{\small Mean time and SD in minutes}}  & \multicolumn{3}{c}{{6.01 (1.65)}} & \multicolumn{3}{c}{{3.80 (0.90)}} & \multicolumn{3}{c}{{13.68 (4.78)}} \\ 
\hline 
 \end{tabular} 

\label{tab:e_err_fini}
\end{table}
\end{landscape}

The results in Table~\ref{tab:e_err_fini} show that the findings from the last section do only partially carry over to the finite sample case. The \ac{bME} approximation shows the smallest \ac{MAE} with \ac{ME} and \ac{SJ} in the second and third place. The major difference between \ac{bME} and \ac{ME} is in the ability to recover the covariance parameters. Both, \ac{ME} and \ac{bME} are slightly superior in recovering the linear parameters when compared to \ac{SJ}.  The \ac{SJ} estimation results showed a greater variability when compared to \ac{ME} and \ac{bME}. The varying ability to recover the linear parameters is clearly indicated by the large \ac{MAE}, especially when compared to the good mean estimate. However, there were, as expected because the estimation was initialized at the true parameter values, no results suspicious of non-convergence.

The results regarding the computation time are of course only informative with respect to the relative difference between the approximations and in this regard they are in line with the previous discussion. Especially \ac{bME} is dramatically slower than \ac{ME} and \ac{SJ}. This has several reasons: First of all this approximation is based on multiple bivariate \ac{MVNCDF}s whose evaluation accounts for most of the computational burden. Second, this leads to -- in terms of computing time -- more costly calculations for the analytic gradient. Third, in order to ensure semi-optimal reordering a full univariate \ac{ME} approximation needs to be computed within the \ac{bME}-algorithm. On the other hand \ac{SJ} is slower than \ac{ME} because of the need to evaluate bivariate \ac{MVNCDF}s and their corresponding gradients.

Table~\ref{tab:e_err_fini_VoT} provides information on the accuracy of estimates of quotients of the 
coefficients such as the value of time in mode choice models. It can be seen that the values for ME and bME 
only slightly decreases while the performance of SJ-1 now is comparable to the one achieved using ME. 
This indicates that the SJ-1 approach shows a larger tendency to converge to various points on the line 
$c \textbf{b}$. Preliminary experiments with random initialization show that this behavior features
very prominently in these cases.

\begin{table}[t]
\centering
\caption{Quotients and corresponding \ac{MAE}s for the linear coefficients across 20 simulated data sets of size 5000 \vspace*{.2cm}}
\begin{tabular} 
 { l c *{2}{c} *{2}{c} *{2}{c}} 
Parameter & true value & \multicolumn{2}{c}{SJ-1} & \multicolumn{2}{c}{ME} & \multicolumn{2}{c}{bME}\\  
&  & {\small mean} &  {\small MAE} & {\small mean}  & {\small MAE} & {\small mean}  & {\small MAE}  \\ \hline 
\multicolumn{8}{l}{{\small linear parameters}}  \\ 
$b_2 / b_1$ & -0.67 & -0.67 & 0.03 & -0.68 & 0.03 & -0.67 & 0.02   \\ 
$b_3 / b_1$ & 1.33 & 1.34 & 0.04 & 1.34 & 0.04 & 1.33 & 0.04   \\ 
$b_4 / b_1$ & 0.67 & 0.66 & 0.02 & 0.66 & 0.02 & 0.66 & 0.02   \\ 
$b_5 / b_1$ & -1.33 & -1.33 & 0.03 & -1.34 & 0.04 & -1.33 & 0.04   \\ 
\hline 
\multicolumn{2}{l}{{\small Mean MAE}}  & \multicolumn{2}{c}{{0.024}} &  \multicolumn{2}{c}{{0.024}}   & \multicolumn{2}{c}{{0.024}}  \\   
\hline 
 \end{tabular} 

\label{tab:e_err_fini_VoT}
\end{table}

It is worth noting that by design the starting values of the optimization are not varied\footnote{The same is true for \citep{bhat2011a} and \citep{cherchi2016}, who all focus on parameter recovery starting from the true values.} but like \ac{MNP} models in general the convergence of an optimizer to a global optimum is not assured for \ac{MaCML} estimation and, therefore, finite-sample results are sensitive to initial values. An investigation of this point is left for further research. 

\begin{table}[t]
\centering
\caption{Mean errors across, sum of absolute differences (SAD) compared to previous estimate and computing time for 10 simulated data sets of size 5000 \vspace*{.2cm}}
\begin{tabular} 
 {l l *{4}{c} } 
 & Tolerance & 0.5$e^{-3}$ & 0.5$e^{-4}$ & 0.5$e^{-5}$  \\ \hline  
SJ-1 & mean MAE & 0.064 &0.097 &0.098  \\ 
& SAD & - &0.422 &0.014  \\  
 & comp. time & 2.25 &7.17 &9.60 \\   
\hline 
ME & MAE & 0.087 &0.088 &0.088  \\ 
& SAD & - &0.022 &0.000  \\  
 & comp. time & 5.49 &6.20 &6.17 \\   
\hline 
bME & MAE & 0.062 &0.063 &0.063  \\ 
& SAD & - &0.020 &0.000  \\  
 & comp. time & 14.49 &22.10 &21.81 \\   
\hline 
 \end{tabular} 

\label{tab:e_err_tol}
\end{table}

In comparing our results with \citep{bhat2011a} we noticed that \citep{bhat2011a} utilize a custom gradient tolerance of $0.5e^{-4}$ in order to assess whether the calculated gradient is numerically zero. 
The lower the gradient tolerance the faster the optimizer terminates once it reaches a point close to the optimum or for that matter any 'flat region' of the pseudo-likelihood. 
The results in Table~\ref{tab:e_err_tol}, which we have generated by simulating 10 data sets of size 5000 for every gradient tolerance, show that lowering the gradient tolerance for our example not only leads to different \ac{MAE}s but also influences the computation times. The main profiteer from a lower gradient tolerance is the \ac{SJ} approximation, while the results of \ac{ME} and \ac{bME} remain remarkably stable. This hints to the fact that estimation performance might be improved by fine-tuning the optimization.

\section{Conclusions} \label{chap:concl}

In this paper we have surveyed several analytic approximations for the \ac{MVNCDF} with
regard to their feasibility in \ac{MNP} model estimation. Using
a simple cross sectional model we compared four approximations from the 
literature (\ac{SJ}-1, \ac{SJ}-A and \ac{ME} as well as \ac{bME}) focusing on estimation rather than prediction accuracy.

Our main results given in section~\ref{chap:large} show that even for the simple model considered in this paper the \ac{MaCML}-estimator is inconsistent for all approximations considered.  The degree of asymptotic bias changes with the specific data generating process but is often very small. We have identified that true 
models with at least one small probability as well as strong correlation structures have an adverse effect 
on estimation accuracy. Furthermore, our simulations show that the parameter estimates are asymptotically more biased than the probability estimates. This is important to keep in mind if the \ac{MaCML}-estimates 
are used to calculate the value of travel time or equivalent quantities involving one or more parameters.  
In general the large-sample results show that \ac{SJ} and \ac{bME} are almost head-to-head with \ac{SJ} 
being superior in special cases while we find that the \ac{ME} approximation championed by \citep{connors2014} is 
dominated by those approximations with respect to asymptotic absolute bias in our setup. 
\\
Still, all approximations 
lead to inconsistent estimators. While the asymptotic bias is small for the example considered there is no guarantee that this is also true for more complicated models. The  results in this paper provide hints on the directions where the search for 'bad' situations continues. 
On the other hand it might be possible to derive explicit bounds for the bias as a function of bounds 
for the approximation error (see e.g. the work of \cite{langdon1984} on the trivariate \ac{ME} approximation). Such a bound would be very helpful as it might allow to identify cases where a large
bias is to be expected and to obtain an (estimated) bound on the magnitude on the bias. This is left 
for future research. 

The results from our finite sample testing, albeit limited to just one
selected model with limited number of replications, show that \ac{bME} and \ac{ME} perform slightly better than \ac{SJ} with only one permutation, which is the default \ac{MaCML} configuration.  This is contrary to the large sample results where the opposite ordering has been found. 
Our finite sample results also  shed some light on the typical runtime of \ac{MaCML} estimation. The computation of \ac{bME} is an order of magnitude slower than both alternatives and hence probably not the preferred option. 
 
Secondly, estimation utilizing \ac{SJ} in the example was slower than \ac{ME}-based \ac{MaCML} estimation due to the 
evaluation of bivariate \ac{MVNCDF}s and its gradients. 
This is especially important in considering internal averaging, 
which is in theory beneficial to \ac{SJ}s performance but adds to the costs from a computational standpoint. 
Jointly the result indicate that the choice of the appropriate approximation to be used is not straightforward and more research is needed. 

The investigation of the approximations demonstrated that the topic of ordering of the  
components for the numerical optimization is a very sensitive one as it has a large impact on the results both in terms of accuracy and in terms of computation time. 
From our results for the limiting pseudo-likelihood for SJ it is demonstrated 
in line with the literature that already using only one random permutation 
delivers similar results to averaging over all combinations. However, it is important to use the same 
ordering for the whole optimization. For ME an optimal ordering might be beneficial but again it is of 
importance that only one ordering is used because otherwise discontinuities in the criterion function 
might cause problems for the numerical optimization. As the optimal ordering depends on the parameter 
values in this case it is unclear how the ordering should be done. Again this is left for future 
research. 

Finally the discussion in section~\ref{chap:finite} suggests that it is warranted to assess the optimization of \ac{MaCML} pseudo-likelihoods more broadly by moving beyond parameter recovery which induces the need to find viable initial estimators. In this respect it is currently totally unknown 
what the consequences of different initialization routines on the properties of the \ac{MaCML} estimators 
are. Preliminary tests suggest that the SJ approach might be more heavily affected than the ME approximations to this issue. Again detailed investigations are left for future research.

\subsection*{Acknowledgments}
This research did not receive any specific grant from funding agencies in the public, commercial, or not-for-profit sectors. The authors would like to thank Daniel Rodenburger for pointing the authors to the idea of unifying the presentation of the \ac{ME} and \ac{SJ} approximation as well as for carrying out some preliminary simulations.

\clearpage
\bibliography{lit}

\clearpage

\end{document}